\documentclass[fleqn,12pt]{article} 
\usepackage[utf8]{inputenc}

\usepackage{authblk}
\usepackage{amsmath}
\usepackage{MnSymbol}%
\usepackage{wasysym}%

\usepackage{placeins} 
\usepackage{pgfplots}
\pgfplotsset{compat=1.14}

\usepackage[algo2e,linesnumbered,ruled,vlined]{algorithm2e}

\usepackage{subcaption}	
\definecolor{dkgreen}{rgb}{0,0.6,0}
\definecolor{gray}{rgb}{0.5,0.5,0.5}
\definecolor{mauve}{rgb}{0.58,0,0.82}
\definecolor{whitegray}{rgb}{0.99,0.99,0.98}
\definecolor{darkgray}{rgb}{0.55,0.55,0.55}
\definecolor{pink}{rgb}{0.9,0.0,0.5}
\definecolor{dkblue}{rgb}{0.0,0.0,0.6}

\usepackage{tikz}
\usetikzlibrary{patterns}
\usetikzlibrary{arrows}
\newcommand{\ie}{\textit{i.e.}}

\definecolor{darkgreen}{rgb}{0.0, 0.2, 0.13}

\SetCommentSty{mycommfont}

\tikzstyle{transition-arrow}=[color=black!50,
        -triangle 90,
        line width=0.5mm,
        postaction={draw, line width=0.2cm, shorten >=0.1cm, -}]
\tikzstyle{hash}=[pattern=north west lines, pattern color=black!50]

  \title{On the improvement of the in-place merge algorithm parallelization}

\author[1,2]{Bérenger Bramas~\footnote{Corresponding author, Berenger.Bramas@inria.fr}}
\author[3,4]{Quentin Bramas}
\affil[1]{CAMUS Team, Inria Nancy - Grand Est, Nancy, France}
\affil[2]{ICPS Team, ICube, Illkirch-Graffenstaden, France}
\affil[3]{Network Team, ICube, Illkirch-Graffenstaden, France}
\affil[4]{University of Strasbourg, France}
  
  \date{March 1, 2020}

  \usepackage{geometry}
      \usepackage{fancyhdr}

\begin{document}
  \pagestyle{fancy}
\renewcommand\headrulewidth{0pt}
\lhead{}\chead{}\rhead{}
\cfoot{\vspace*{1.5\baselineskip}\thepage}

  \flushbottom
\maketitle
\thispagestyle{empty}

  \begin{abstract}
  {In this paper, we present several improvements in the parallelization of the in-place merge algorithm, which merges two contiguous sorted arrays into one with an $\mathcal{O}(T)$ space complexity (where $T$ is the number of threads).
  The approach divides the two arrays into as many pairs of partitions as there are threads available; such that each thread can later merge a pair of partitions independently of the others.
  We extend the existing method by proposing a new algorithm to find the median of two partitions.
  Additionally, we provide a new strategy to divide the input arrays where we minimize the data movement, but at the cost of making this stage sequential.
  Finally, we provide the so-called linear shifting algorithm that swaps two partitions in-place with contiguous data access.
  We emphasize that our approach is straightforward to implement and that it can also be used for external (out of place) merging.
  The results demonstrate that it provides a significant speedup compared to sequential executions, when the size of the arrays is greater than a thousand elements.}
  \end{abstract}

  
  \section{Introduction}
  
  The problem of in-place merging is defined as follows.
  Having two sorted arrays, $A$ and $B$, of sizes $L_A$ and $L_B$ respectively, we want to obtain a sorted array $C$ of size $L_A+L_B$, which contains $A$ and $B$.
  A naive approach to solve this problem is to have $C$ as an external buffer of size $L_A+L_B$ and to construct it without modifying the sources $A$ and $B$.
  However, having an in-place algorithm to perform this operation could be necessary if the amount of available memory is drastically limited relatively to the data size, or could be potentially faster especially if the allocation/deallocation is costly or the given arrays are tiny.
  
  In-place merging, but also sorting based on in-place merge, have been studied for a long time.
  A so-called fast in-place merging has been proposed~\cite{bib:praticalmerging} and is the most known strategy as it has a linear time complexity regarding the number of elements, which makes it the optimal solution.
  The main idea of this algorithm is to split the partitions into blocks of size $\sqrt{L_A+L_B}$ and to use one block as a buffer.
  One of the main assets of the method is that working inside a block with quadratic complexity algorithms will be of linear complexity relatively to ${L_A+L_B}$.
  Variations of this algorithm have been proposed~\cite{bib:todo3,bib:todo6,bib:todo5} where the authors provide an important theoretical study.  
  Shuffle-based algorithms have been proposed~\cite{bib:todo2} where the elements of the $A$ and $B$ are first shuffled before being somehow sorted.
  
  Additionally, the community has provided several in-place merge sorts~\cite{bib:todo1,bib:todo7,bib:todo4} that either use an algorithm similar to the fast in-place merging or they do not do it in-place by using more than $\mathcal{O}(1)$ space, or even use a classical sort algorithm to merge the partitions resulting in a $\mathcal{O}(N \, log(N) )$ complexity, where $N$ is the number of elements.  
  Parallel merge sort has also been investigated~\cite{7036012} and is usually implemented by, first, sorting independent parts of the array in parallel, and then using a single thread to merge each pair of previously sorted partitions.
  Parallel out of place merge of two partitions has also been proposed~\cite{green2014merge}.
  
  Odeh et al.~\cite{6270834} proposed a simple algorithm to find the optimal partition points when merging two arrays.
  They also provide advanced cache optimizations and obtained significant speedup.
  However, their approach is not in-place. 
  
  The possibly closest related work to our paper is the parallel sort algorithm proposed by Akl et al.~\cite{5009478}. It is based on a parallel in-place merging strategy that is very similar to our approach.
  They also use an algorithm to find a median value to divide one pair of partitions into two pairs of partitions (from two to four, from four to eight, and so on), such that working on each pair can be done in parallel.
  In the current work, we extend their strategy to find the median, and we concentrate on the parallelization of the merge, leaving aside the fact that it can be used to sort in parallel.

  The  contributions of this study are the following:
  \begin{itemize}
  \setlength\itemsep{0em}
  \item Provide a new strategy to find the median in a pair of partitions (where the two partitions have potentially different sizes);
  \item Provide a new partition exchange algorithm called linear shifting;
  \item Provide a new method to split the original partitions with the minimum data movement.
        Note that the method can only be used if the elements to merge can be used to store a marker;
  \item Study the complexity of our method, and prove the correctness of the circular shifting;
  \item Describe the full algorithms, giving details on the corresponding implementation~\footnote{The complete source code is available at https://gitlab.inria.fr/bramas/inplace-merge . The repository also includes the scripts to execute the same benchmarks as ones presented in this paper and the results.} and providing an extensive performance study.
  \end{itemize}
  
  The paper is organized as follows.
  In Section~\ref{sec:background}, we provide a brief background related to in-place merging, partitions swapping and divide and conquer parallelization.
  Then, in Section~\ref{sec:mergeingalgorithm}, we describe our parallel strategies and the linear shifting method.
  We provide the details regarding the complexity of the algorithm in Section~\ref{sec:complexities}.
  Finally, we provide the performance results in Section~\ref{sec:performancestudy}.
  
  
  \section{Background}
  \label{sec:background}

\paragraph{Notations}
  Consider that $A$ and $B$ are two arrays, $A<B$ means that the elements of $A$ are smaller than the elements of $B$ ($>$, $\leq$ and $\geq$ are defined similarly). $|A|$ denotes the size of $A$, and $A[a:b]$ denotes the slice of $A$ starting from the element of index $a$ to the element of index $b$.
  
  \paragraph{Merging in-place}
  
  The so-called fast in-place merging algorithm~\cite{bib:praticalmerging} is the reference for merging in-place as it can do it with linear complexity.
  The algorithm splits the two input arrays into blocks of size $\sqrt{N}$, where $N$ is the total number of elements, and uses one of the blocks as a buffer.
  The algorithm starts by moving the greatest elements of the input, taken from the ends of $A$ and $B$, in the first block that become the buffer, in front of the input.
  Then, the algorithm moves the other blocks so that they are lexicographically ordered.
  Swapping two blocks takes $\mathcal{O}(\sqrt(N))$ operations but the blocks are already almost sorted so in this step has a $\mathcal{O}(N)$ complexity.
  Then, the smallest element at the right of the buffer is moved to the left the buffer.
  Doing so, the buffer traverses the input array from left to right and when it reaches the end of the array, all the elements on its left are sorted.
  Finally, the buffer is sorted, which terminates the merge.
  
  \paragraph{Circular shifting}
  
\begin{figure}[ht!]
  \centering
  \includegraphics[width=\textwidth, height=.1\textheight, keepaspectratio]{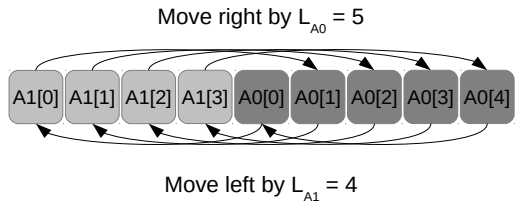}
  \caption{Exchange/swapping of the two partitions $A_1$ and $A_0$ using the circular shifting algorithm.
           In this case, only one cycle is needed.
           The algorithm can be implemented without temporary using only a swap function between items.}
  \label{fig:./Fig1}
  \end{figure}

  A different in-place merging algorithm was proposed with an algorithmic complexity of $\mathcal{O}(L_A^2+L_B)$~\cite{bib:circularshifting}.
  The quadratic coefficient of the complexity comes from the fact that the first partition of length $L_A$ can potentially be shifted $L_A$ times.
  The algorithm performs the following steps.
  It starts by skipping the elements of $A$ that are smaller than $B[0]$ because they are already at the correct position: we reach a new initial configuration with smaller arrays.
  Then, when the case $A[0] > B[0]$ is met, the algorithm moves the elements $B[0:K]$ lower than $A[0]$ to their correct position by swapping them with the $K$ first elements of $A$.
  It gives the configuration $B_0 \, A_1 \, A_0 \, B_1$, where $A_0$ and $B_0$ are of length $K$, $A_0 \leq A_1$, $B_0 < B_1$, and the elements in $B_0$ are at the correct position.
  The algorithm continues by exchanging partitions $A_1$ and $A_0$ to reach again an initial configuration with two partitions to merge: $swap(A_1 , A_0) \, B_1 \rightarrow A \, B_1$.
  To swap the two partitions, the authors proposed a method called circular shifting, which is an in-place exchange with the minimum data movements.
  It is known where each element should go: the elements in $A_1$ have to be shifted to the right by $Length(A_0)$, while the elements in $A_0$ have to be shifted to the left by $Length(A_1)$.
  Therefore, the circular shifting starts with the first element and moves it to the right while putting the element to be overwritten in a buffer.
  Then it moves the element in the buffer to its correct position by performing a swap.
  The algorithm continues until the element in the buffer is the first element that has been moved, i.e. a cycle is performed when the iteration is back to the first element, as it is illustrated by Figure~\ref{fig:./Fig1}.
  Several cycles can be needed to complete the entire partition swapping (more precisely $LCM(L_A,L_B)$\footnote{$LCM(a,b)$ denotes the least common multiple of $a$ and $b$} cycles are needed).
  Despite its efficiency in terms of complexity and data movements, this algorithm suffers from a poor data locality because it has irregular memory accesses, and goes forward and backward.
  More details concerning the circular shifting are given in Section~\ref{sec:inplacerearrange}.
  
  \paragraph{Parallelization of divide and conquer strategies}
  
  The parallelization of divide and conquer algorithms have been previously done by the community.
  Among them, the Quick-sort have been particularly studied~\cite{tsigas2003simple,Evans1985,1675993,blelloch1996programming}.
  This algorithm sorts an array by dividing it into two partitions, where the first one includes the smallest elements and the second one the largest elements.
  Then, the same process is recursively applied until all partitions are of size one, and the array sorted.
  A straightforward parallelization consists of creating one task for each of the recursive calls and to place a synchronization afterward.
  As possible optimizations, one can limit the number of tasks by stopping the creation of tasks at a certain recursion depth, and to use a different but faster sort algorithm when the partitions are small enough.

  \paragraph{Parallel in-place merge and median search}  
  
  Akl et al.~\cite{5009478} have proposed a parallel merge using a divide and conquer strategy that they used to parallelize a sort algorithm.
  In the first stage, the master thread finds the lower median of partitions $A$ and $B$ (which is the median of $C$ without actually forming it).
  Then, the master thread delegates the elements of $A$ and $B$ greater than the median to another process, and focuses on the elements lower than the median.
  After $T$ subdivisions, each thread can merge a pair of partitions without any memory conflict.
  The proposed algorithm is not in-place, and consequently, there is no discussion about the swapping of partitions.
  As a remark, we were not able to find the optimal median with the method proposed by the authors in their study and it is not clear to us if the method they describe only works if both partitions are of the same size.
    
  
  \section{In-place merging parallelization}
  \label{sec:mergeingalgorithm}
    
\begin{figure}[ht!]
  \centering
  \includegraphics[width=\textwidth, height=.25\textheight, keepaspectratio]{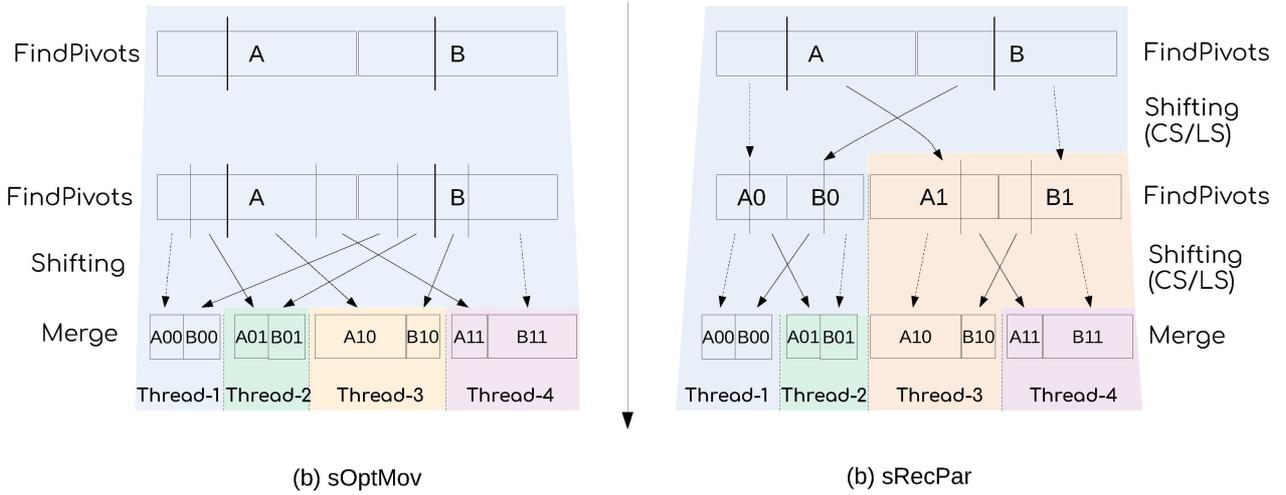}
  \caption{Example to merge two partitions using the sRecPar (a) and sOptMov (b) strategies with 4 threads.
          (a) With sOptMov, the master thread finds all the pivots one level of division after the other.
          Then, it shifts all the partitions directly at the right position.
          Finally, the pairs of partitions are merged by all the threads.
          (b) With sRecPar, the master thread finds the pivots and shift the two center partitions using either the circular shifting (CS) or the linear shifting (LS).
          Then, it creates a task to delegate the work on $A_1$ and $B_1$.
          The same technique is applied by both threads until the desired number of pair of partitions is reached.
          Finally, the pairs of partitions are merged by all the threads.}
  \label{fig:example}
  \end{figure}

  The main idea consists in splitting the first two initial partitions into as many partitions as the number of threads, and to merge these sub-partitions in parallel independently one from another.
  More formally, we start with the transformation $ A \, B \rightarrow A_0 B_0 A_1 B_1 ... A_{T-1} B_{T-1}$, with $X_i < Y_{i+1}$, where $X, Y \in [A,B]$ and $0 \leq i < T$.
  To obtain this configuration, we propose two different methods.
  In the first one, we first find the splitting intervals and build the partitions with the minimum number of data move (algorithm~sOptMov).
  In the second, we recursively sub-divide two partitions into four (two pairs of partitions) that can be processed in parallel (algorithm~sRecPar).
  This second method shares many similarities with the algorithm from Akl et al.
  In both cases, we use a method call $FindMedian$ to find where two partitions should be split and subdivided into fours.
  After the division stage, each thread of id $i$ merges $A_i$ and $B_i$.
  Figure~\ref{fig:example} contains an illustration of the two methods using four threads.
  
  \subsection{Finding the median to divide two partitions into four ($FindMedian$)}
  
  Considering that two partitions $A$ and $B$ have to be split, the $FindMedian$ algorithm finds a pivot value and two indexes, one for both partitions, such that splitting the partitions $A$ and $B$ to obtain four partitions $A_0$, $A_1$ and $B_0$ $B_1$ using the indexes will respect the following properties:
  $A_0 \leq A_1$ and $B_0 \leq B_1$ by definition, and $A_0 \leq B_1$ and $B_0 \leq A_1$ by construction.
  To do this, we need to find a pivot value $p_v$ such that $A_0 \leq p_v \leq A_1$ and $B_0 \leq p_v \leq B_1$, but we also want to obtain balanced partitions such that $|A_0| + |B_0| \approx |A_1| + |B_1|$.
  Finding the optimal median can be done with a $\mathcal{O}(L_A \cdot log(L_B) )$ complexity by iterating over the elements of $A$ and consider each of them as the pivot and then find the position of the corresponding pivot in $B$ using binary search.
   
  Instead, we propose to use a double binary search on both partitions.
  We first put two indexes $p_a$ and $p_b$ in the middle of $A$ and $B$, respectively, and we compare several criteria.
  For instance, we compare the elements $A[p_a]$ and $B[p_b]$, but also the sizes of the partitions before and after the pivots to decide how these should be moved with the objective of reducing the difference between $|A_0| + |B_0|$ and $|A_1| + |B_1|$.
  For example, if $A[p_A] < B[p_B]$, we have to move $p_a$ to a greater element (increasing the index) or move $p_B$ to a lower element (decreasing the index).
  Considering that $|A_0|+|B_0| < |A_1|+|B_1|$, we move $p_a$, hence we increase $|A_0|$ (and decrease $|A_1|$).
  We continue the process until we cannot go deeper, i.e.\ at least one of the search intervals is of size 1.
  The complexity of the algorithm is $\mathcal{O}(log(L_A) + log(L_B))$.
  The Algorithm~\ref{algo:FindMedian} provides all complete details.
  Note that we test at the beginning of the algorithm if $A < B$ or $A > B$ to treat these two cases separately.
  This can be seen as a specialization of the method because we know that the values returned by $FindMedian$ are going to be used to swap the partitions, and thus we are returning indexes that will reduce the workload in the final merge algorithm.
  
    \begin{algorithm2e}
    \DontPrintSemicolon 
    \KwIn{$A$ and $B$ the partitions to split}
    \KwOut{$p_A$ and $p_B$ the indexes of the pivots in $A$ and $B$ respectively}
    \SetKwProg{Fn}{Function}{}{}
        \Fn{FindMedian(A,B)}{
            \If{$|A| = 0$ \textbf{OR} $|B| = 0$ \textbf{OR} $A[last] \leq B[0]$}{
                return $|A|$,0\;
            }
            \If{NOT $A[0] \leq B[last]$}{
                return 0,$|B|$\;
            }
            $left_A \gets 0$\;
            $limit_A \gets |A|$\;
            
            $left_B \gets 0$\;
            $limit_B \gets |B|$\;
            
            \While{$left_A < limit_A$ \textbf{AND} $left_B < limit_B$ \textbf{AND} $A[p_A] \neq B[p_B]$}{
                $p_A \gets (limit_A - left_A)/2 + left_A$\;
                $p_B \gets (limit_B - left_B)/2 + left_B$\;
                $A0 \gets p_A$\;
                $A1 \gets |A| - p_A$\;
                $B0 \gets p_B$\;
                $B1 \gets |B| - p_B$\;
                \If{$A[p_A] < B[p_B]$}{
                    \If{$A0+B0 < A1+B1$}{
                        $left_A \gets p_A + 1$\;
                    }
                    \Else{
                        $limit_B \gets p_B$\;
                    }
                }
                \Else{
                    \If{$A0+B0 < A1+B1$}{
                        $left_B \gets p_B + 1$\;
                    }
                    \Else{
                        $limit_A \gets p_A$\;
                    }
                
                }
            }
            $p_A \gets (limit_A - left_A)/2 + left_A$\;
            $p_B \gets (limit_B - left_B)/2 + left_B$\;
            
            \Return{\{$p_A$,$p_B$\}}\;
        }
    \caption{{\sc FindMedian}}
    \label{algo:FindMedian}
    \end{algorithm2e}
  
  \subsection{Division of the arrays with the sOptMov strategy}
  
  The sOptMov strategy starts by finding all the pivots, without moving any data, and then moves the data directly in the right position.
  In this end, the algorithm calls the $FindMedian$ function $T-1$ times, leading to a $\mathcal{O}(T \cdot (log(L_A) + log(L_B)) )$ complexity.
  Once we know the interval/size of each sub-partition, we can compute where each element has to be moved, and we can move them directly to the right position with a $\mathcal{O}(1)$ space complexity.
  However, we must allocate an array of intervals of size $T$ that will contain the original positions of the partitions and their destination positions.
  By doing so, we can know in $\mathcal{O}(log(T))$ where an element at position $i$ should be moved by finding its corresponding partition in the array.
  The move algorithm performs as follow.
  It iterates over the elements of the array and tests each of them to know if it has already been moved or if it is at the right position.
  If the element has not been moved, it performs a move cycle that starts from this element using the array of intervals to know where to move the elements.
  Such a move cycle is similar to a move cycle in the circular shifting algorithm, except that we have to use the array of intervals to know where each element has to be moved.
  Consequently, the resulting complexity to move the element is $\mathcal{O}(log(T) \cdot (L_A + L_B) )$.
  
  To mark which elements have been moved, it would seem natural to use a secondary array of Boolean, but this is impossible in our case because we want to have a $\mathcal{O}(1)$ space complexity.
  In this aim, we propose a technique where we store a marker directly in the elements.
  As a consequence, the sOptMov remains in-place if and only if the data type that is merged can store a marker.  
  For instance, consider that the elements to merge are integers.
  We can find a marker value by making the difference between the greatest and lowest value in the array $M = 1 + MaxVal - MinVal$, where $MaxVal = Max(array[:])$ and $MinVal = Min(array[:])$.
  To mark an element, we simply add $M$ to it, and to test if a value is marked, we test if it is greater than $MaxVal$.
  The method is valid if $MaxVal + M$ does not overflow, or more precisely, if $M + MaxVal - MinVal$ does not overflow if we scale the values using $MinVal$.
  After the data has been moved, the values should be unmarked by subtracting $M$.
  
  After that the master thread has partitioned the array into $T$ pairs of partitions, it creates one parallel task per pair of partitions and finally waits for the completion of the tasks.
  
  Algorithm~\ref{algo:parallel_merge_soptmove} provides a simplified implementation of sOptMove.
  The algorithm starts by evaluating the number of recursive levels that are needed depending on the number of threads at line~\ref{algo:parallel_merge_opt_soptmove:depth}.
  Then, the master thread finds all the intervals of values that are moved in-place line~\ref{algo:parallel_merge_opt_soptmove:reorder}.
  Finally, a parallel section is created, and each thread merges two sub-partitions into one line~\ref{algo:parallel_merge_soptmove:merge}.
  
    \begin{algorithm2e}
    \DontPrintSemicolon 
    \KwIn{$array$ an array of $size$ elements sorted between 0 and $middle$, and $middle$ and $size$}
    \KwOut{an array that contains the same elements but which is fully sorted}
    \SetKwProg{Fn}{Function}{}{}
     \Fn{sOptMove\_parallel\_merge(array, middle, size)}{
        \If{$middle = 0$ \textbf{OR} $middle = size$ \textbf{OR} $array[middle-1] \leq array[middle]$}{
            return\;
        }
        \tcc{Number division levels}
        $depthLimit \gets ffs(nbThreads) - 1$\; \label{algo:parallel_merge_opt_soptmove:depth}
        
        \tcc{Build an interval structure using the parameters}
        $intervals[0][0] \gets init\_intervals([0, middle], [middle, size])$\;
        
        \For{idxDepth from 0 to depthlimit-1}{
             \tcc{Divide partions from one level to the other}
             $nbSplits \gets pow(1, idxDepth)$\;
             \For{idxSplit from 0 to nbSplits-1}{
                 $middleA, middleB \gets$ FindMedianFromInterval($intervals[idxDepth][idxSplit], array$)\;
                 
                $intervals[idxDepth+1][idxSplit \times 2] \gets$ left\_interval($currentInterval, middleA, middleB$)\;
                $intervals[idxDepth+1][idxSplit \times 2+1] \gets$ right\_interval($currentInterval, middleA, middleB$)\;
             }
        }
        
        \tcc{Build the partitions from the intervals}
        $finalPositions \gets$ reorder\_multi($array, intervals[depthLimit][:]$)\; \label{algo:parallel_merge_opt_soptmove:reorder}
        
        \textcolor{blue}{\#pragma omp parallel}\;
        \For{idxPartition from 0 to nbThreads-1}{
            \tcc{Each thread merges two sub-partitions together}
            $partitionPosition \gets finalPositions[idxPartition]$\;
            std::inplace\_merge($array + partitionPosition.start, array+partitionPosition.middle, array+partitionPosition.end$)\; \label{algo:parallel_merge_soptmove:merge} 
        }
    }
    \caption{{\sc Parallel Merge - sOptMove}}
    \label{algo:parallel_merge_soptmove}
    \end{algorithm2e}
    
  \subsection{Division of the arrays with the sRecPar strategy}  
  
  In this strategy, we recursively partition the arrays and create a task for each recursive call.
  In more details, the master thread finds the pivots to partition $A$ and $B$, and use them to swap the partitions and obtained $A_0 B_0 A_1 B_1$.
  Then, the same principle should be applied on $A_0 B_0$ and $A_1 B_1$ separately.
  Therefore, the master thread creates a task to delegate the work on $A_1 B_1$, and continues the process on $A_0 B_0$.
  The final partitions and intervals are the same as the ones in sOptMov, but in sRecPar the partitions are moved as soon as possible and in parallel.
  A consequence is that a part of the elements is moved multiple times.
  
  Algorithm~\ref{algo:parallel_merge} provides the details of the sRecPar strategy.
  First, we compute the depth of the recursive calls at line~\ref{algo:parallel_merge:depth}.
  We want to have one task per thread; hence the depth is equal to $log2(T)$ (considering that the number of threads is a power of two).
  As long as the size of partitions is greater than a threshold limit and that the depth level is not reached (line~\ref{algo:parallel_merge:while}), we call the $FindMedian$ function to know where to split the partitions (line~\ref{algo:parallel_merge:findmiddle}), and then we shift the partitions to obtain four of them (line~\ref{algo:parallel_merge:shift}).
  We create a task to have a thread working on the last two partitions (line~\ref{algo:parallel_merge:omp}).
  Note that we create only one task because there is no need to create a second task and a second recursive call: the current thread is the one that should process the first two partitions and creating a second task would imply significant overhead.
  The current thread merges its two partitions line~\ref{algo:parallel_merge:merge}, and then waits for the completion of the tasks it has created.
    
    A critical operation of this algorithm is the shifting procedure that is used to swap/permute the center partitions.
    This operation can be implemented in different manners.
    
    \begin{algorithm2e}
    \DontPrintSemicolon 
    \KwIn{$array$ an array of $size$ elements sorted between 0 and $middle$, and $middle$ and $size$}
    \KwOut{an array that contains the same elements but which is fully sorted}
    \SetKwProg{Fn}{Function}{}{}
        \Fn{sRecPar\_parallel\_merge(array, middle, size)}{
            \If{$middle = 0$ \textbf{OR} $middle = size$ \textbf{OR} $array[middle-1] \leq array[middle]$}{
                return\;
            }
            
            \textcolor{blue}{\#pragma omp parallel}\;
            \textcolor{blue}{\#pragma omp master}\;
            \{\;
                \tcc{Number division/recursion levels}
                \qquad $depthLimit \gets ffs(nbThreads) - 1$\; \label{algo:parallel_merge:depth}
                \qquad sRecPar\_parallel\_merge\_core($array, 0, middle, size, 0, depthLimit$)\;
            \}\;
        }
        \Fn{sRecPar\_parallel\_merge\_core(array, left, middle, size, depth, depthLimit)}{ 
            \tcc{If the partitions are not empty}          
            \If{$currentStart \neq currentMiddle$ \textbf{AND} $currentMiddle \neq currentEnd$}{
                \tcc{Division until depthLimit}
                \While{$level \neq depthLimit$ \textbf{AND} $(currentEnd-currentStart) > SIZE\_LIMIT$}{  \label{algo:parallel_merge:while}     
                    \tcc{Find where to divide the current partitions}
                    $middleA, middleB \gets$ FindMedian($\&array[currentStart],  currentMiddle - currentStart, \label{algo:parallel_merge:findmiddle} 
                                                   \&array[currentMiddle],  currentEnd-currentMiddle$)\;
                                          
                    $sizeRestA \gets currentMiddle-currentStart-middleA$\;
                    $sizeRestB \gets currentEnd-currentMiddle-middleB$\;
                                
                    \tcc{Shift the values using circular or linear shifting}
                    shift($array + middleA + currentStart, sizeRestA, middleB+sizeRestA$)\; \label{algo:parallel_merge:shift} 
                    
                    \tcc{Create a task to have thread working on the righ sub-partitions}
                    \textcolor{blue}{\#pragma omp task default($none$) firstprivate($array, level, currentStart,$ $ currentMiddle, currentEnd, middleA, middleB$)}\; \label{algo:parallel_merge:omp} 
                    parallel\_merge\_core($array,
                                             currentStart+middleA+middleB,
                                             currentStart+middleA+middleB+sizeRestA,
                                             currentEnd,
                                             level+1, depthLimit$)\;
                    
                    $currentEnd \gets currentStart+middleA+middleB$\;
                    $currentMiddle \gets currentStart+middleA$\;       
                              
                    $level \gets level + 1$\;
                }

                \tcc{Merge the remaining partitions}
                std::inplace\_merge($array + currentStart, array+currentMiddle, array+currentEnd$)\; \label{algo:parallel_merge:merge} 
        
                \textcolor{blue}{\#pragma omp taskwait}\;
            }
        }
    \caption{{\sc Parallel Merge - sRecPar}}
    \label{algo:parallel_merge}
    \end{algorithm2e}

  
  \subsection{Shifting algorithms (in-place permutation)}
  \label{sec:inplacerearrange}
  
  As input, the shifting algorithm takes an array composed of two partitions, $A$ and $B$ of lengths $L_A$ and $L_B$, respectively, and aims to inverse the partitions: the elements of $A$ should be moved to the right by $L_B$, and the elements of $B$ to the left by $L_A$.  
    
  
  \paragraph*{Linear shifting (LS)}
  
\begin{figure*}[ht!]
  \centering
  \includegraphics[width=0.7\textwidth]{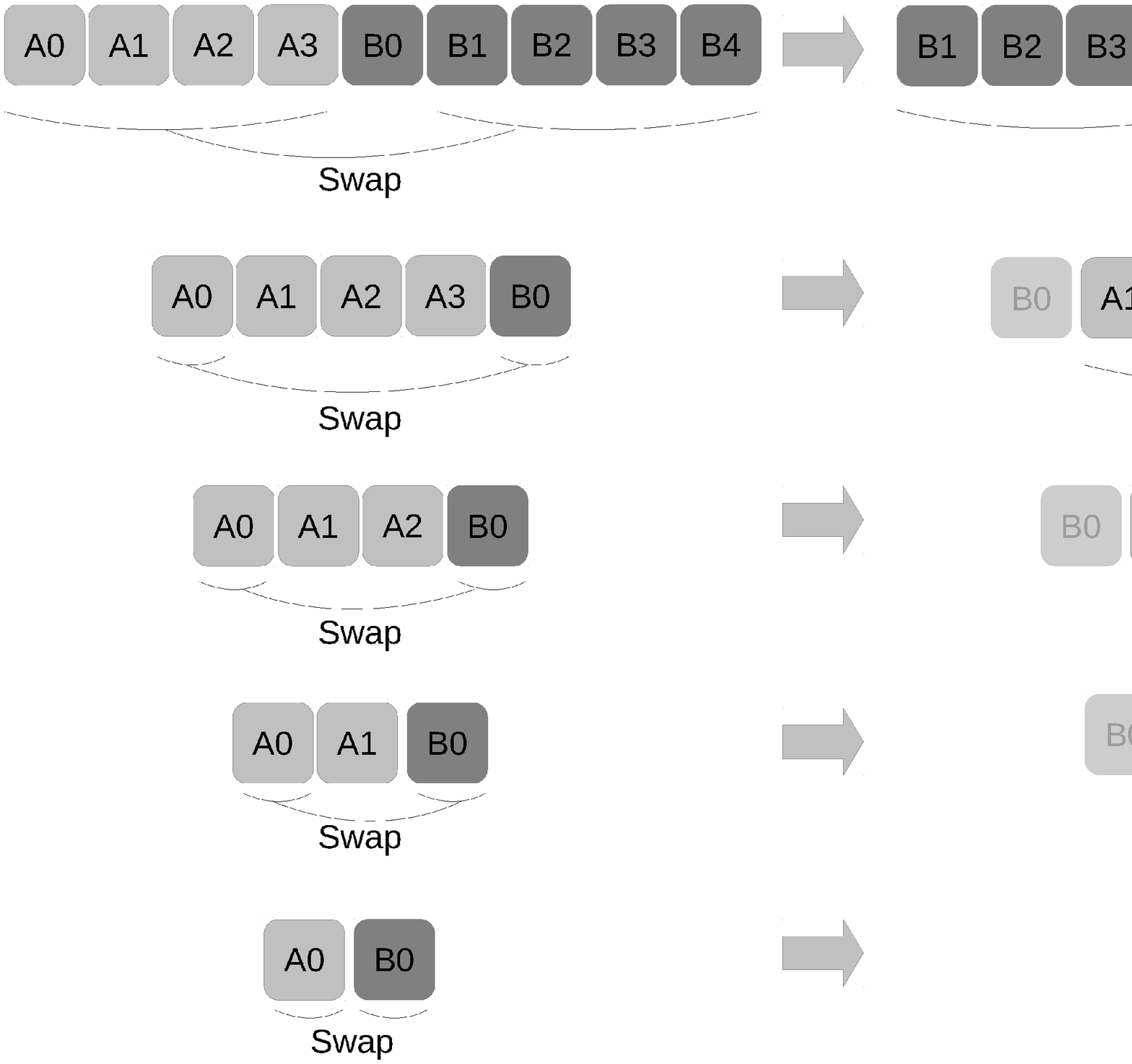}
  \caption{Linear shifting example. 
           The smallest partition on the left is moved by swapping elements with the largest partition.
           Then, the smallest partition is at the correct position (light gray on the right).
           The process is applied until all elements have been moved at their correct positions.
           Note that the notation is updated at each row to use $A$ and $B$.}
  \label{fig:./Fig3}
  \end{figure*}

  The algorithm starts by moving the smallest partition, let us say $A$, to the appropriate position.
  In this end, it swaps all the elements of $A$ with those that are currently there, which belong to $B$, ending in the configuration $B_1 B_0 A$.
  Then the smallest partition $A$ will remain untouched because it has been moved to the correct place.
  We end up with the same problem but with different data; the largest partition has been split into two parts, the part that has been moved, and the part that has not changed $B_1 B_0 \rightarrow A B$.
  Therefore, we can use the algorithm again.
  Despite its simplicity and the fact that it does not have to be implemented recursively, this algorithm can have a significant overhead as it potentially moves some of the data multiple times.
  On the other hand, it uses regular and contiguous memory access, which can make it extremely efficient on modern CPU architectures.
  Figure~\ref{fig:./Fig3} presents several iterations of this method.
   
  \paragraph*{Circular shifting (CS)}
  
  We describe the circular shifting to make the paper self-content, but we remind that this algorithm has been proposed by Dudzińki et al~\cite{bib:circularshifting}.
  When we have to swap two partitions, we already know where each element of the input array should be moved.
  The elements of $A$ should be shifted to the right by $L_B$, whereas the elements of $B$, should be shifted to the left by $L_A$.  
  The Figure~\ref{fig:./Fig1} shows the principle of the circular shifting.
  
  If both partitions have the same length, then we can swap the $i$\textsuperscript{th} element of $A$ and of $B$, for all $i$, $0\leq i < L_A = L_B$.
  
  If one of the two partitions is longer than the other, we will shift some elements inside the source partition itself.
  For example, if $L_A < L_B$, all of the elements of $A$ will go to the position originally taken by $B$, whereas some elements of $B$
  will go to the previous part of $A$ and some others on the previous part of $B$.
  This comes from the interval relation: $B$ is located in the input array in the indexes interval $[L_A,L_I-1]$, resp. in the output array in the indexes interval $[0,L_B-1]$, and $[L_A,L_I-1]\cap [0,L_B-1] = [L_A, L_B - 1]\neq \emptyset$.
  
\begin{figure}[ht!]
      \centering
      \includegraphics[width=.6\textwidth, height=.5\textheight, keepaspectratio]{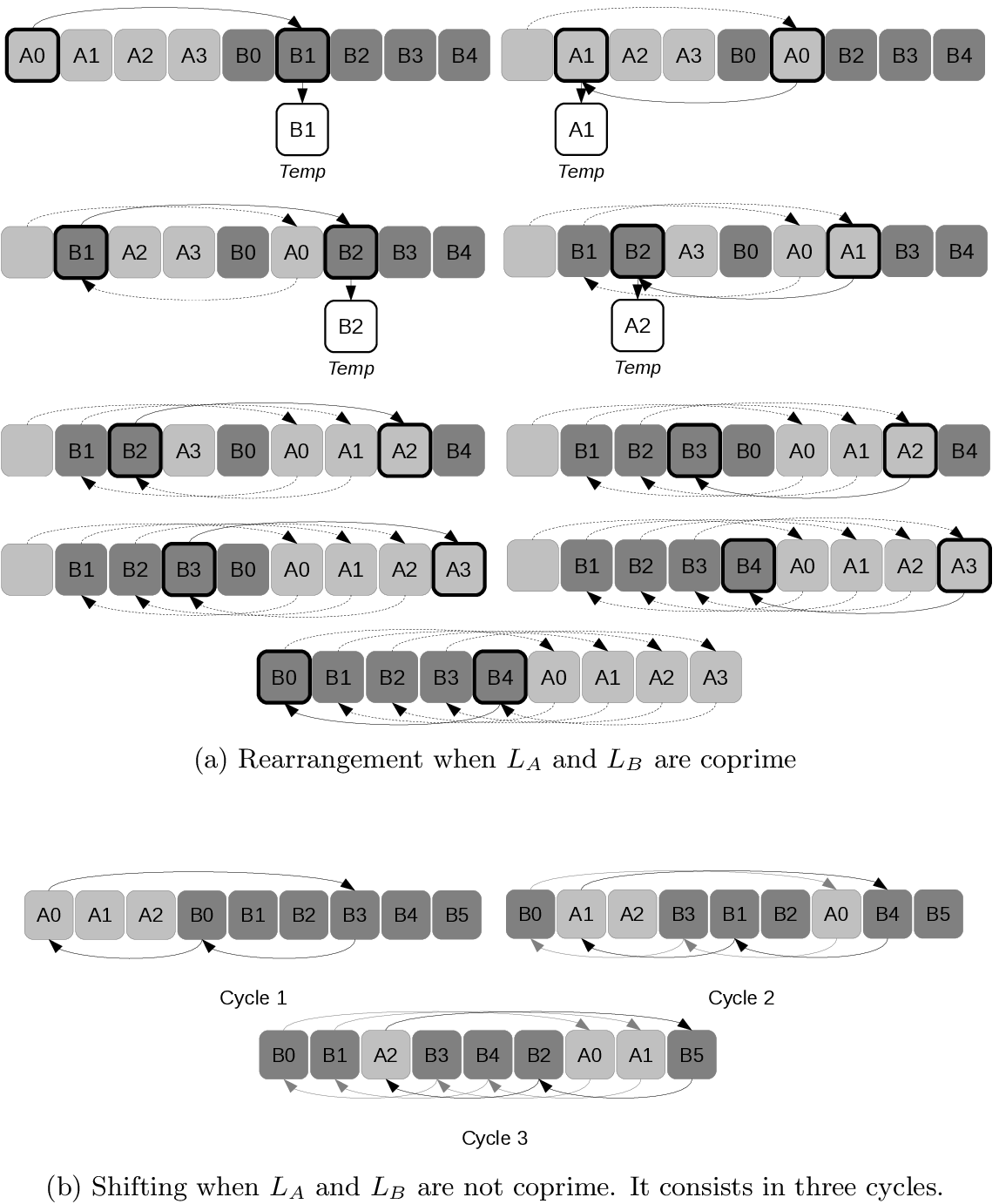}
      \caption{Circular shifting examples.}
      \label{fig:rearrangefullmod}
  \end{figure}

  From this relation, we can move the elements, one after the other.
  Considering that $Input$ is the array that contains the elements of $A$ and of $B$.
  Starting from the first element $Input[0]$, we shift it to the right, by $L_B$, $Input[0] \rightarrow Input[L_B]$.
  In order not to lose data, we save the destination element $Input(L_B)$ into a temporary buffer.
  Then, either the saved element has to be shifted by $-L_A$ (if it belongs to $B$) or shifted by $L_B$ (if it belongs to $A$).
  After shifting the saved element, we replace the temporary buffer by the new destination and repeat the operation, as illustrated in Figure~\ref{fig:rearrangefullmod}(a).
  
  We prove below that the sequence of indexes forms a cycle and goes back to the first position of the loop after several iterations.
  Then, we can replace the element in the first position with the one in the temporary buffer, since it had been moved at the first iteration.   
    
  When the first cycle stops, the process is not over because some elements may not have been moved yet.   
  We need to perform additional cycles starting from different indexes, as illustrated in Figure~\ref{fig:rearrangefullmod}(b).
        
  To guarantee that this algorithm is completed, we need to ensure that all of the elements have been moved and moved only once.
  
  \paragraph*{Circular shifting proof of correctness}
   
  We consider the case where $L_A < L_B$ but the inverse can be proved similarly.
  To understand the circular shifting process one can follow these steps:
   \begin{enumerate}
   \item We start from index $i_0$, $0\leq i_0 < L_A$.
   \item Recursively, for all $k\geq 0$, if $i_k < L_A$, then the destination index of the element at index $i_k$ is the $i_{k+1} = i_k + L_B$ (the element at $i_k$ is shifted by $L_B$ to the right).
   Otherwise, if $i_k \geq L_A$, then the destination index of the element at index $i_k$ is the $i_{k+1} = i_k - L_A$.
   \end{enumerate}
   
   Since the number of indexes is finite, there are two integers $h$ and $l$, $h < l$ such that $i_h = i_l$. One can see that the first times this occurs we have $h=0$. Indeed, if $h > 0$ is the first such integer, then $i_{h-1}\neq i_{l-1}$, and $i_h$ is the destination index of the element at index $i_{h-1}$ and at index  $i_{l-1}$ which contradicts the fact that there is a bijection between the elements of the $Input$ and the $Output$ arrays.
   Hence, we have $i_0 = i_l$ and 
   \begin{align}
   & i_l = i_0 + x_1*L_B - x_2*L_A \\
   & x_1*L_B = x_2*L_A
   \end{align}
   with $x_1$ and $x_2$ coprime.
   The solution of this equation is 
   \[x_1 = \frac{LCM(L_A, L_B)}{L_B}\text{ and }x_2 = \frac{LCM(L_A, L_B)}{L_A}\]
   where $LCM$ is the least common multiple (LCM).
   
   For instance, if $L_A$ and $L_B$ are coprime, then $x_1 = L_A$ and $x_2 = L_B$, which means that during the first cycle, the $L_A$ elements from $A$ have been shifted by $L_B$ positions to the right and the $L_B$ elements from $B$ have been shifted by $L_A$ to the left before reaching back the first position \ie, all elements are moved when the cycle is finished.

   If $L_A$ and $L_B$ are not coprime, we have $x_1 \ne L_A$ and $x_2 \ne L_B$.
   In this case, only $x_1+x_2$ elements have been moved to their correct position during the first cycle.
   To complete the algorithm, we need to perform more cycles but starting at different positions.
   Since $i_0$ was arbitrary in the above analysis, we observe that each cycle moves the same number of elements: 
   \begin{align*}
    x_1+x_2 
    &= \frac{LCM(L_A, L_B)(L_A+L_B)}{L_AL_B} \\
    &= \frac{L_A+L_B}{GCD(L_A, L_B)} 
\end{align*}
   Where $GCD(a,b)$ denotes the greatest common divisor of $a$ and $b$.
   Moreover, one can see that the sets of indexes visited by two cycles are either identical or disjoint.
   Indeed, if an index $i_k$ is visited by two cycles starting at $i_0$ and at $i_0'$, then $i_{k+1}$ is also visited by the two cycles.
   Finally, $L_A$ and $L_B$ are coprime if and only if there exists $y_1$ and $y_2$ such that $y_1*L_B - y_2*L_A = 1$ \ie, the index $i_0 + 1$ is reached by a cycle if and only if $L_A$ and $L_B$ are coprime.
   Hence, if $L_A$ and $L_B$ are not coprime, then $i_0+1$ is not visited by the cycle starting at $i_0$ and we can start the next cycle from $i_0+1$ to move $x_1+x_2$ more elements. To verify the condition that $i_0 < L_A$ we can choose $i_0+1 \mod L_A$ and the same remains true, or we can simply start the first cycle at $i_0 = 0$.
   
   By the previous analysis, we see that, in order to move all the elements at their correct position, we have to perform $t$ cycles starting at indexes $0$, $1$, $\ldots$, $t-1$, with $t = (L_A + L_B) / (x_1 + x_2) = GCD(L_A, L_B)$, which corresponds effectively to the smallest positive number $t$ such that $y_1*L_B - y_2*L_A = t$. 
  
  
  \subsection{Complexities}
  \label{sec:complexities}
  
  \paragraph*{Linear shifting}
  
  When swapping two partitions $A$ and $B$ of sizes $L_A$ and $L_B$, each iteration consists in swapping all the elements of the smallest partition to their correct position.
  If $L_A = L_B$, each swap locates two elements to their correct positions, so there are exactly $L_A$ swaps.
  Otherwise, each swap locates only one element to its correct position.
  The resulting complexity is $\mathcal{O}(L_A + L_B)$, but there are up to $2 \times (L_A + L_B)$ swaps.
  The memory access pattern is contiguous (linear and regular).
  
  \paragraph*{Circular shifting}
  
  In this case, each element is read once and moved directly to its correct position.
  We have $t = GCD(L_A,L_B)$ cycles, and one cycle moves $(L_A+L_B)/t$ elements.
  So, the complexity is also linear, and there are exactly $L_A+L_B$ swaps, but the memory access pattern is irregular.
  In fact, from one cycle to the next one, the spatial difference is only one, but inside a cycle, each loop may access different parts of the array.  
  
  \paragraph*{In-place merging}

  If we use the state-of-the-art implementation, then the complexity of merging two partitions in-place in linear.  
  
  \paragraph*{Parallel in-place merging}
  
  With the sOptMov strategy, finding the spiting positions and shifting the partitions are two different operations of complexities $\mathcal{O}(T \cdot (log(L_A) + log(L_B)))$ and $\mathcal{O}(log(T) \cdot (L_A + L_B)$, respectively.
  With the sRecPar strategy, the master thread is the one that perform the more divisions in the creation of the partitions, and the complexity of its work is $\mathcal{O}(log(T) \times (log(L_A) + log(L_B) + L_A + L_B))$.
  Therefore, the difference between both methods is an order of magnitude $T$.  
  In both strategies, each thread merges $(L_A + L_B)/T$ elements on average.
  If we consider that the merge of two partitions is linear and that the number of threads is a constant, the resulting complexity is linear $\mathcal{O}(log(L_A) + log(L_B) + L_A + L_B)$.

  
  \section{Performance Study}
  \label{sec:performancestudy}
  
  \paragraph{Hardware}
  The tests have been done on an Intel Xeon Gold 6148 (2,4GHz) Skylake with 20 cores and 48GB of memory.
  The cache sizes are L1 32K, L2 1024K and L3 28160K.
  The system also has 48GB additional memory connected to another CPU within the same node.
  
  \paragraph{Software}
  We used the GNU compiler Gcc 8.2.0 and executed the tests by pinning each thread on a single core (\emph{OMP\_PLACES=cores} \emph{OMP\_NUM\_THREADS=T} \emph{OMP\_PROC\_BIND=TRUE} \\ \emph{OMP\_WAIT\_POLICY=ACTIVE}).
  We use up to 16 threads for the parallel versions.
  
  \paragraph{Test cases}
  To evaluate the performance of our methods, we measured the time taken to merge two arrays.
  We tested with arrays of different sizes, from $S=2^2$ to $S=2^{22}$, and for elements of different  sizes from 4 to 65540 bytes (note that the size of the array was limited when using elements of 65540 bytes due to main memory capacity). The first 4 bytes are used when comparing elements pairewise.
  The memory occupancy of an array is given by $S \times element\_size$.
  We split the two partitions at different positions located at $1/4$, $1/2$ and $3/4$ of the whole array, and the elements of each partition are generated using the formula $array[i] = rand\_0\_1() \times 5 + array[i-1]$, with $array[0] = 0$.
  We compared our methods with the fast in-place merging algorithm~\cite{bib:praticalmerging} and with the classic merge with external buffer, where we included the time needed for the allocation, the copy, and the deallocation.
  Each dot in our graphs are obtained by averaging 50 runs.
  
  We use the C++ standard std::inplace\_merge function as a reference, but also in our parallel implementation, i.e.\ each thread uses this function to merge its partitions.
  
\begin{figure*}[ht!]
      \centering
      \includegraphics[width=\textwidth, height=\textheight, keepaspectratio]{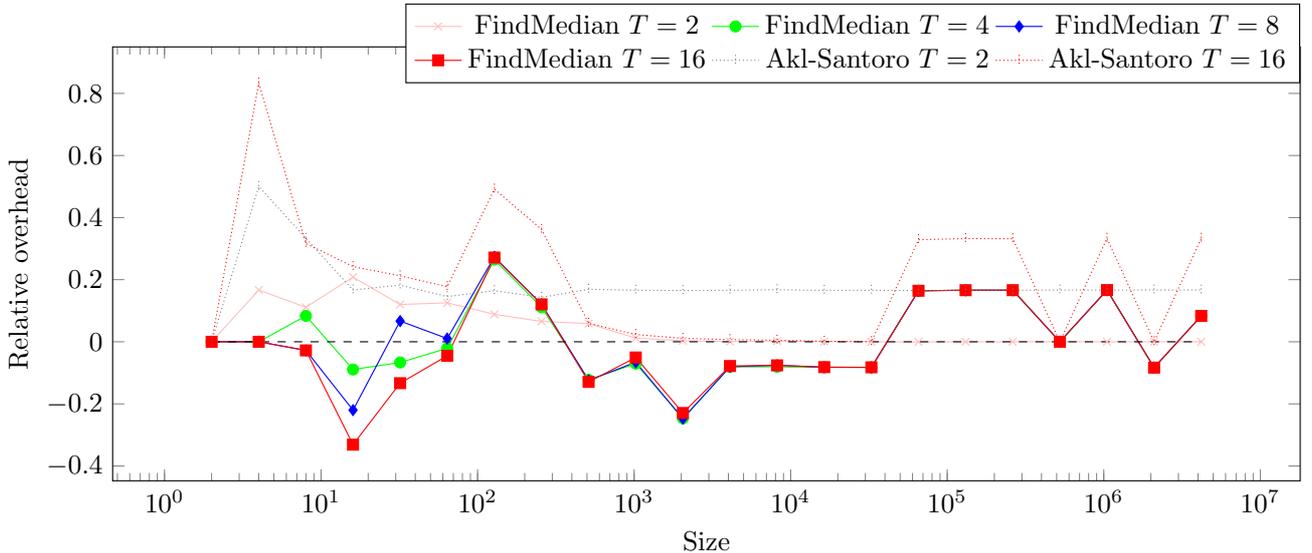}
    \caption{Relative partition size differences between the $FindMedian$ and the optimal search of pivots.
            The difference is obtained by $(Max_{FindMedian}-Max_{opti})/Max_{opti}$, where $Max_{FindMedian}$ is the size of the largest pair of partitions when using $FindMedian$, and $Max_{opti}$ is the size of the largest pair of partitions when using the optimal method.
            The parameter $T$ represent the number of splits of the input array.
            The same comparison is performed for the algorithm proposed by Akl et al. (Akl-Santoro).}
    \label{fig:diffopti} 
\end{figure*}

  \paragraph{Results} 
  In Figure~\ref{fig:diffopti}, we provide the theoretical difference between our $FindMedian$ heuristic and the optimal search algorithm.
  We can see that, for the test cases we use, the difference is not significant.
  Surprisingly, using the $FindMedian$ method can provide better results when there is more than one level of division ($T > 2$).
  This means that the optimal division at the first level is not necessarily optimal when we have to perform multiple divisions.
  Hence, using the double binary search provides intervals that are as good as the ones obtained with the optimal method in most cases.
  We also put the results we obtained with the algorithm by Akl et al.
  It appears that for the first level of division ($T=2$), it does not find the optimal median and add a significant overhead.
  For more than one level of division ($T > 2$), the overhead decreases but remains important.
  
\begin{figure*}[ht!]
      \centering
      \includegraphics[width=\textwidth, height=\textheight, keepaspectratio]{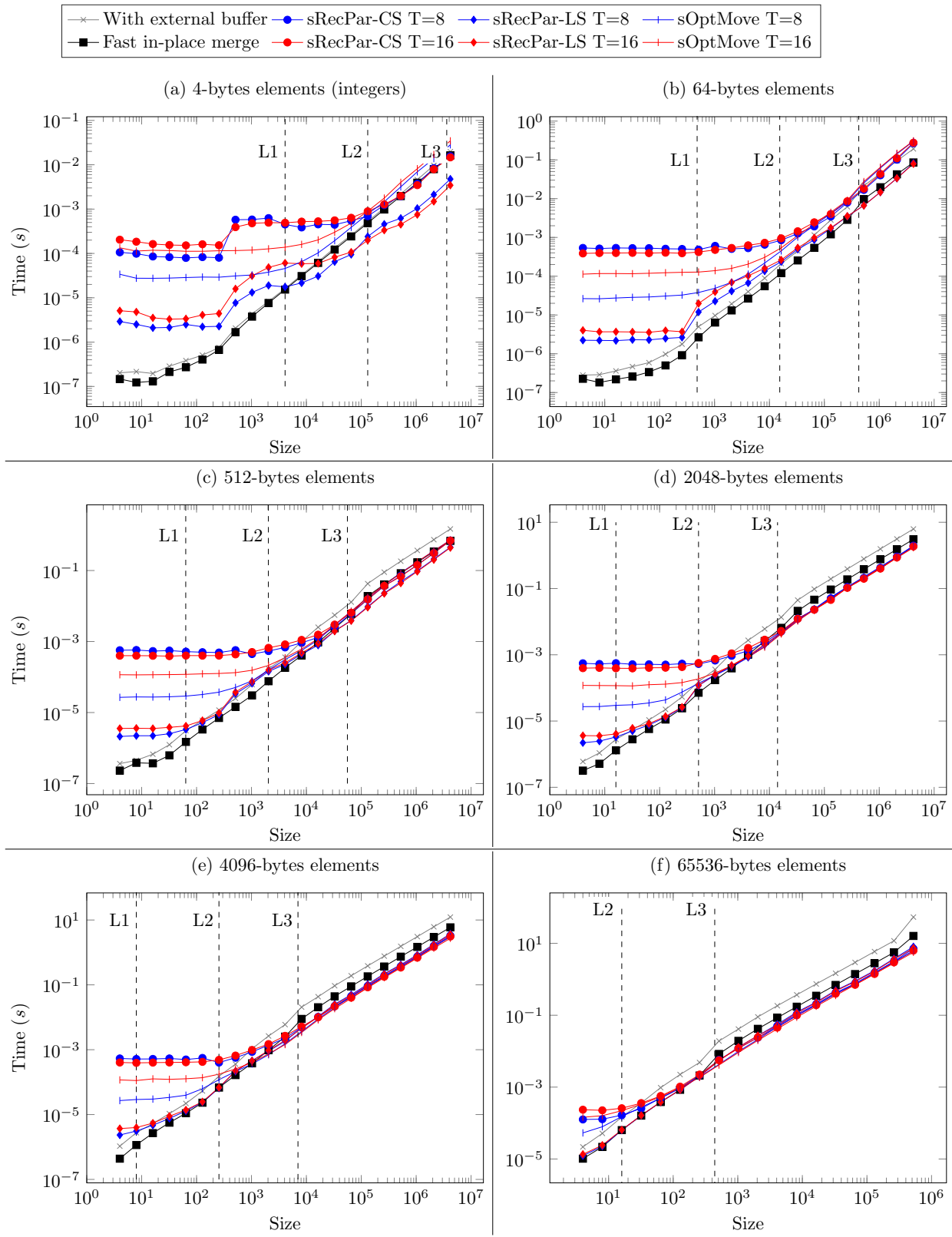}

    \caption{Execution time to merge two partitions store in an array of \emph{Size} elements.
            The time is obtained by testing three partition configurations $1/4$, $1/2$, $3/4$ of random values within the same range, and performing $50$ merge to show only the average.
            LS and CS acronyms mean linear and circular shifting, respectively.
            The vertical bars represent the different CPU caches regarding the size of the items.}
    \label{fig:timing} 
\end{figure*}

\begin{figure*}[ht!]
      \centering
      \includegraphics[width=\textwidth, height=\textheight, keepaspectratio]{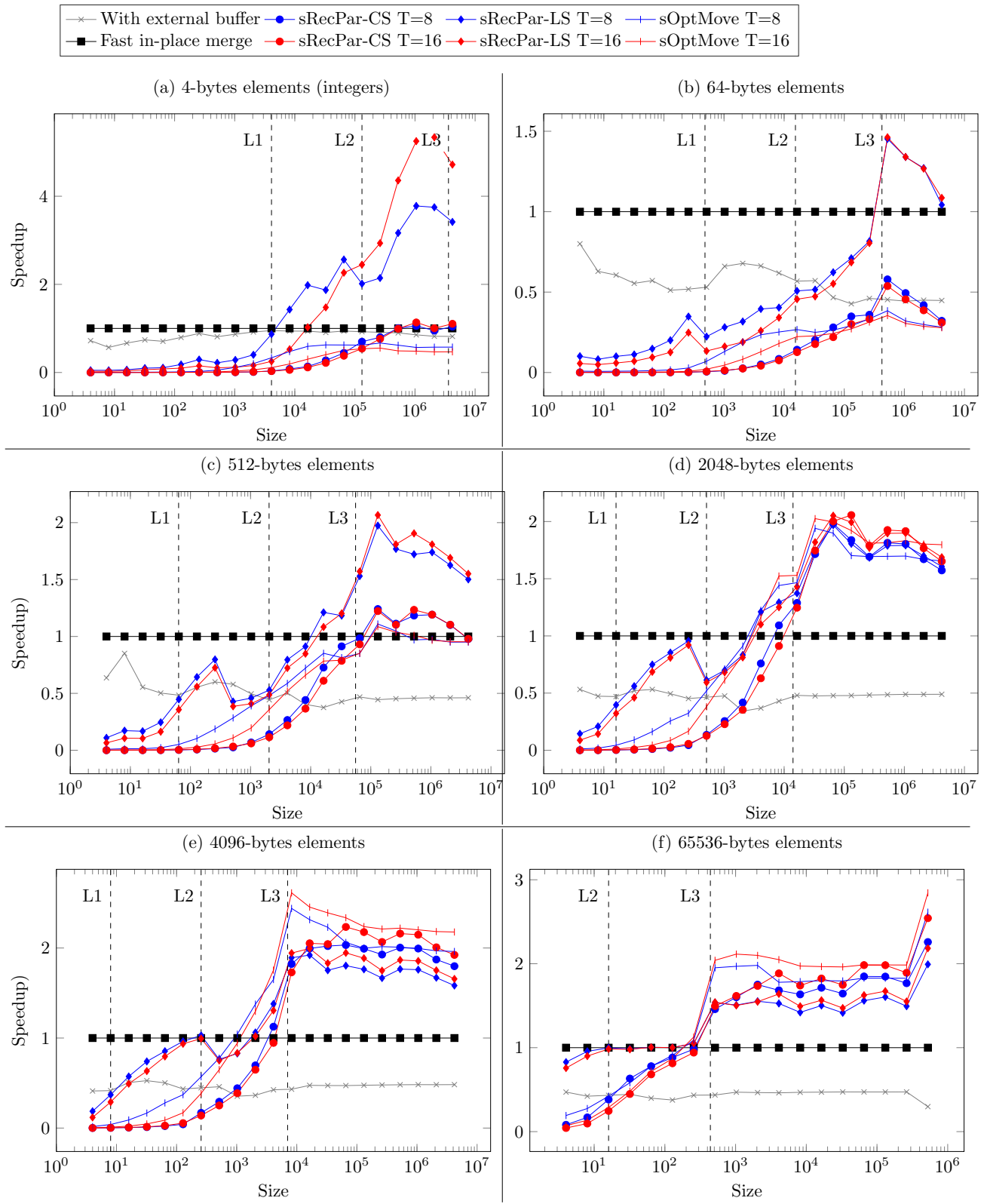}

    \caption{Speedup of the different methods vs. the sequential fast in-place (std::inplace\_merge) when merging two partitions store in an array of \emph{Size} elements.
            The execution times used to compute the speedup are obtained by testing three partition configurations $1/4$, $1/2$, $3/4$ of random values within the same range, and performing $50$ merge to show only the average. 
            LS and CS acronyms mean linear and circular shifting, respectively.
            The vertical bars represent the different CPU caches regarding the size of the items.}
    \label{fig:speedup} 
\end{figure*}

  In Figure~\ref{fig:timing}, we provide the execution times for the different arrays and element sizes, and we provide the same results by showing the speedup in Figure~\ref{fig:speedup}.
    
  The merge with external buffer (\textcolor{gray}{$\times$}) is not competitive against the fast in-place merge (\textcolor{black}{$\blacksquare$}).
  Looking at the execution time, the method appears to be always slower, and this is indeed obvious when we look at the speedup (Figure~\ref{fig:speedup}).
  The method seems not sensitive to the fitting of the data in the caches, but we can see that the larger the data the slower is the method compared to the fast in-place merge.
  Therefore, \emph{(i)} the allocation of a new array, \emph{(ii)} the fact that this new memory block is not in the cache initially, and \emph{(iii)} that both arrays may compete to stay in the caches,  add a significant overhead, which motivates the use of in-place merging.
  
  If we compare the three parallel methods sRecPar-LS ($\blacklozenge$), sOptMove ($|$) and sRecPar-CS ($\bullet$), we see that they are significantly impacted by the fitting of the data in the caches.
  For instance, if we look at the execution time, Figure~\ref{fig:timing}, as long as the data fit in the L2 cache the best performance are obtained by sRecPar-LS ($\blacklozenge$), then sOptMove ($|$) and finally sRecPar-CS ($\bullet$) no matter the number of threads (\textcolor{blue}{8}/\textcolor{red}{16}) or the size of the elements (from Figures (a) to (f)).
  We even see a steady state, and sRecPar-LS takes more or less the same duration to merge 4 elements or the maximum number of elements that fit in L1.
  Similarly, sRecPar-CS takes the same duration to merge 4 elements or the maximum number of elements that fit in L2.
  However, if we look at the details for the array of integers, Figure~\ref{fig:timing}(a), we see a jump between $10^2$ and $10^3$ elements for sRecPar-CS, even if $10^3$ should fit in the L1 cache.
  The only difference between sRecPar-LS and sRecPar-CS is the shifting method used to swap the partitions in the division stage.
  Consequently, we see the effect of having irregular memory accesses (jumps and in backward/forward directions) when we use the CS method.
  From the execution time, when the data does not fit in the L2 cache, we can notice that all the methods appear to have a linear complexity (each line has a unique slope).
  
  If we now look at the parallel methods compared with the fast in-place merge (\textcolor{black}{$\blacksquare$}) in terms of speedup, Figure~\ref{fig:speedup}, we see that the parallel methods are never faster than the fast in-place merge if the data fit in the L1 cache.
  In these configurations, the execution time is between $10^{-7}s$ and $10^{-6}s$, which make it almost impossible to be executed in parallel faster than in sequential.
  Moreover, our parallel versions have an additional stage - the division of the input array - that the sequential method avoids.
  Then for the merge of arrays of integers, Figure~\ref{fig:speedup}(a), the sRecPar-LS ($\blacklozenge$) provides a speedup when the data does not fit in the L1 cache.
  For all the other sizes, a parallel method provides a speedup starting when the data almost does not fit in L3.
  When the elements are small, Figure~\ref{fig:speedup} (a), (b) and (c), sRecPar-LS ($\blacklozenge$) is superior.
  However, when the elements become larger, the methods are equivalent for 512 integers, and then the order is inverse for elements of sizes 1024 and 16384: sOptMove ($|$) is faster than sRecPar-CS ($\bullet$), which is itself faster than sRecPar-LS ($\blacklozenge$).
  This makes sense, because sOptMove ($|$) will have the minimum number of data displacement, and sRecPar-CS ($\bullet$) uses circular shifting in the division stage, which is optimal to swap partitions.
  Consequently, the best method depends on the size of the elements.
  
  Using 16 (\textcolor{red}{-}) threads instead of 8 (\textcolor{blue}{-}) provides a speedup in most cases, but there is a clear drop in the parallel efficiency as the speedup is much lower than a factor 2.
  This means that the overhead of dividing the array increases with the number of threads, while the benefit is attenuated.
  
  We notice a jump in the speedup for size $2^{22}$ and element size 16384 (Figure~\ref{fig:speedup}(f)), but from the execution time (Figure~\ref{fig:timing}(f)), we can see that the parallel versions are stable, but the sequential method slowdown.
  We suspect that NUMA effects are then mitigated thanks to the use of more cache memory in the parallel versions (since each core has its L2 and L1 cache).

  
  \section{Conclusion}
  \label{sec:conclusion}
  
  We have proposed new methods to parallelize the in-place merge algorithm.
  The sOptMove approach splits the input arrays sequentially but with the minimal number of moves, whereas the sRecPar approach splits the input arrays in parallel but at the cost of extra moves.
  Both methods rely on double binary search heuristic to find the median with the aim of dividing two partitions into four.
  Additionally, we propose a linear shifting algorithm that swaps two partitions with contiguous memory accesses, which appears to be significantly more efficient than the circular shifting.
  The spacial complexity of our parallel merge is $\mathcal{O}(T)$ and the algorithm complexity is linear if the core merge algorithm used by the thread is also linear.
  From our performance study, we show that our approaches are competitive compared to the sequential version and allow to obtain a significant speedup on arrays of large elements that do not fit in L3 cache or for arrays of integers that do not fit in the L1 cache.
  We also demonstrate that our double binary search heuristic is close to the optimal find median search when the array contains regular increasing values of the same scale.
  
  As a perspective, we would like to create a similar implementation on GPUs.
  This will require the evaluation and adaption of the shifting strategies to find the most suitable strategy for this architecture.
  We will also have to find a method to ensure that more than one thread group performs the initial partitioning.
  
  \section*{Acknowledgments}
  
  Experiments presented in this paper were carried out using the Cobra cluster from the Max Planck Computing and Data Facility (MPCDF).
  
  \FloatBarrier
  
  \bibliographystyle{unsrt} 
  \bibliography{Inplace-merge-algorithm}%

  \FloatBarrier
  
  
   
\end{document}